# Research article

# Development of a fully deep learning model to improve the reproducibility of sector classification systems for predicting unerupted maxillary canine likelihood of impaction.

**M. Galdi**, DDS, Department of Medicine, Surgery and Dentistry, University of Salerno, Via S. Allende, 84081 Baronissi, SA, Italy; email: magaldi@unisa.it

**D. Cannatà**, DDS, Department of Medicine, Surgery and Dentistry, University of Salerno, Via S. Allende, 84081 Baronissi, SA, Italy; email: dcannata@unisa.it

**F. Celentano,** DDS, Department of Medicine, Surgery and Dentistry, University of Salerno, Via S. Allende, 84081 Baronissi, SA, Italy; email: f.celentano14@studenti.unisa.it

**L. Rizzo**, PhD student, Department of Computer Science, University of Salerno, Via G. Paolo II, 84084 Fisciano, SA, Italy; email: lrizzo@unisa.it

**D. Rossi**, P PhD student, Department of Computer Science, University of Salerno, Via G. Paolo II, 84084 Fisciano, SA, Italy; email: dorossi@unisa.it

**T. Bocchino**, Adjunct Professor, Department of Neuroscience, Reproductive Science and Dentistry, University of Naples Federico II, 80131 Naples, NA, Italy; email: tecla.bocchino2011@gmail.com

**S. Martina**, Associate Professor, Department of Medicine, Surgery and Dentistry, University of Salerno, Via S. Allende, 84081 Baronissi, SA, Italy; email: smartina@unisa.it

**Abstract**

**Objectives.** The aim of the present study was to develop a fully deep learning model to reduce the intra- and inter-operator reproducibility of sector classification systems for predicting unerupted maxillary canine likelihood of impaction.

**Methods.** Three orthodontists (Os) and three general dental practitioners (GDPs) classified the position of unerupted maxillary canines on 306 radiographs (T0) according to the three different sector classification systems (5-, 4-, and 3-sector classification system). The assessment was repeated after four weeks (T1). Intra- and inter-observer agreement were evaluated with Cohen's K and Fleiss K, and between group differences with a z-test. The same radiographs were tested on different artificial intelligence (AI) models, pre-trained on an extended dataset of 1,222 radiographs. The best-performing model was identified based on its sensitivity and precision.

**Results.** The 3-sector system was found to be the classification method with highest reproducibility, with an agreement (Cohen's K values) between observations (T0 versus T1) for each examiner ranged from 0.80 to 0.92, and an overall agreement of 0.85 [95% confidence interval (CI) = 0.83-0.87]. The overall inter-observer agreement (Fleiss' K) ranged from 0.69 to 0.7. The educational background did not affect either intra- or inter-observer agreement (p>0.05). DenseNet121 proved to be the best-performing model in allocating impacted canines in the three different classes, with an overall accuracy of 76.8%.

**Conclusion.** AI models can be designed to automatically classify the position of unerupted maxillary canines.

# 1. Introduction

The management of impacted maxillary canines, whose reported prevalence ranges between 0.8 and 4.7% [1–4], presents a challenge to clinicians due to their impact on facial aesthetics and oral function [5]. Indeed, maxillary canines support the upper lip and alar base, contributing significantly to facial esthetics, and their proper size, shape, and alignment enhance the attractiveness of the smile; moreover, they functionally help disclosing the posterior teeth during lateral jaw movements [6].

A prompt interception of impacted maxillary canine in early mixed dentition has been reported to reduce the risk of potential implications of impaction, such as root resorption of neighboring teeth, and the need for surgical exposure of the impacted canines [7,8]. Therefore, the early management of unerupted canines with unfavorable eruption pattern potentially streamline or erase the complexity of more extensive therapy in the future, minimizing overall treatment time and cost [9].

Some clinical signs have been identified as predictors of an unfavorable canines' eruption path, including the delayed eruption of the permanent canine, the prolonged retention of the primary canine beyond the age of 14–15 years, the absence of a labial canine bulge, the presence of a palatal prominence, and associated dental anomalies such as distal tipping or ectopic positioning of the lateral incisor [10,11]. Moreover, several radiographic features in panoramic radiographs (PRs) have been associated with unerupted maxillary canine likelihood of impaction, notably the angulation between the long axis of the impacted canine and the midline, the vertical distance from the canine cusp tip to the occlusal plane, and the sector location of the canine's medial crown position relative to the lateral incisor [12]. Accordingly, Ericson and Kurol segmented the unerupted canine's surrounding area into 5 distinct sectors based on its relationship to the adjacent lateral incisor. Each sector was associated with a more or less favorable prognosis of canine eruption [13,14]. Subsequently, simplifications of the 5-sector classification have been proposed by several authors, but the limitations related to the time and experience required to perform manual classification persist [15].

Artificial intelligence (AI) methods, including deep learning (DL), are increasingly employed in dentistry for a wide range of analytical purposes, such as landmark localization, image segmentation, and object detection, progressively allowing the automation of diagnostic workflows and clinical decision-making processes [16,17].

A recent comprehensive review reported that current AI technologies can substantially assist dental professionals in the interpretation of PRs, achieving high accuracy in detecting carious lesions, osteoporosis, maxillary sinusitis, periodontal bone loss, periapical pathologies, and in tooth identification [18]. Nevertheless, among the potential applications of AI in orthodontics, the issue of impacted canines is not included. Indeed, although automating the classification of canine impaction on PRs could enhance clinical decision-making and reduce both the time and effort required by practitioners, the intricate morphology of dentofacial structures and the superimposition of anatomical boundaries on PRs present considerable difficulties for AI algorithms in recognizing maxillary canine impactions [19]. Therefore, at present, no fully automated, end-to-end system exists for the classification and treatment planning of impacted canines.

As initial stage for developing comprehensive fully automated tools to support clinical decision-making pertaining to impacted maxillary canines, AI-driven platforms that successfully detect impacted and nonimpacted maxillary canines on cropped and uncropped PRs were demonstrated [20,21]. The present study aimed to take a further step forward in this process, developing a method to automatically classify the position of unerupted maxillary canines according to the sector classification systems (SCSs) described in literature.

## 2. Methods

### 2.1 Ethical approval and consent

This study received approval from the Campania Region 2 Ethics committee (protocol code: 2024/30960). Patients (or parents/guardians if a minor) consented with written consent from the first visit to the screening and triage clinic at the Dental Complex Operating Unit of Dentistry of the Hospital "San Giovanni di Dio e Ruggi d'Aragona" in Salerno. The written consent indicated that data and radiographs would be used for research and education purposes without identification. After receiving ethical approval on 06 December 2024, the Dental Complex Operating Unit obtained access to data collection from the medical reports and radiographs of the patients. Permission to access the data for retrospective collection of information started on 20 December 2024.

### 2.2 Study sample and dataset

A retrospective analysis of 17854 PRs was conducted using data from the Dental Complex Operating Unit of Dentistry of the Hospital "San Giovanni di Dio e Ruggi d'Aragona" in Salerno, taken between 2013 and 2024. PRs were screened by two dentists (LR and SM) according to the following criteria.

Inclusion criteria:
- PRs with high quality and good visibility of anatomic structures;
- presence of monolateral (either right or left) or bilateral unerupted maxillary canines;

- patients older than 9;
- no gender restriction.

Exclusion criteria:

- PRs with poor quality or reduced visibility of anatomic structures;
- pathological situations (endocrinal deficiency such as hypothyroidism, hypopituitarism, trauma, or jaw fractures), and hereditary diseases or syndromes such as Down's syndrome and cleidocranial dysostosis affecting facial or dental growth pattern;
- absence of lateral incisors adjacent to the impacted canine;
- dental anomalies of the upper incisors;
- presence of a neoformation (cyst, odontoma) in the upper jaw.

When PRs displayed bilateral unerupted canines, it was split in two radiographs (A and B), separately considered in the analysis.

## 2.3 Classifications

The SCSs used in the present analysis were described in Table 1 and Figure 1.

*Figure 1*. Sector classification system according to Ericson and Kurol (a), Lindauer (b), Kim (c).

## 2.4 Examiners and calibration

Six practitioners randomly selected among two groups of clinicians with different expertise in orthodontics (orthodontic specialists versus general dental practitioners), each including six clinicians of the Department of Medicine, Surgery, and Dentistry "Scuola Medica Salernitana", University of Salerno, Italy, were invited to classify unerupted canines featured in PRs according to the three SCSs. Before the assessment, all the examiners attended a one-hour training session, during which an expert in sector classification (SM) educated them in identifying sectors and determining the position of canines on a set of PRs with unerupted canines.

After the training session, random 20% sample of digital radiograph of unerupted maxillary canines taken from the larger set of digital radiographs were given to the examiners, who were invited to locate the medial crown position of the unerupted canine in one of the five sectors identified by Ericson and Kurol [13], in one of the four sectors identified by Lindauer [22], and in one of the three sectors described by Kim [23]. Each examiner was invited to perform sector classification in a quiet room, without time limit (time T0). After a 4-week interval, all the examiners were invited to repeat the sector classification with the same digital radiographs placed in a different order (time T1).

## 2.5 Statistical Analysis

To estimate Cohen's kappa and Fleiss' kappa, a sample size of 96 subjects was needed, based on a 95% confidence interval (CI), a margin of error of 0.2, an expected positive test rate of approximately 20%, and without making any assumptions about the true kappa value. The training was evaluated by computing the Cohen's kappa between the examiners and the trainer (SM) at the different observations (T0 and T1). The range of variation of K statistic is between 0 for no agreement and 1 for perfect agreement with five intermediate levels: *'slight agreement'* (0.01–0.20), *'fair agreement'* (0.21–0.40), *'moderate agreement'* (0.41–0.60), *'substantial agreement'* (0.61-0.80) and *'almost perfect agreement'* (0.81–0.99) [24]. Cohen's kappa value and 95% CIs were calculated for single examiners and overall to evaluate the intra-examiner agreement between observations (T0 vs. T1). The agreement between observations (T0 vs. T1) was also computed for each group (Os vs. GDPs) and between group differences in sector classification agreement were assessed through z-Test. Fleiss's Kappa was used to assess inter-examiners agreement, both in aggregate and divided into groups, at T0 and T1. Between groups differences in intra-examiner agreement were assessed through a z-Test. Intra- and Inter-examiners agreement were calculated for all SCSs. The statistical significance was set at p-value < 0.05. SPSS software version 30.0 (SPSS Inc., Chicago, IL, USA) was used.

## 2.6 Deep Learning Model Development

### 2.6.1 Sector Classification and Data Encoding

The model was built upon the SCS that demonstrated the highest level of agreement between observers in the preliminary phases of the study.

For supervised training, each sector was converted into a numerical format using one-hot encoding, allowing the model to distinguish between multiple output categories.

Images were resized to a standardized shape and normalized before being used for training. Clinical data were prepared separately and included as complementary input.

### 2.6.2 Model Architecture

To improve computational efficiency while preserving predictive performance, a knowledge distillation framework was employed to train the AI model. This method involves training a smaller, simplified neural network, referred to as the *student model*, to reproduce the behavior of a larger and more expressive *teacher model*.

The teacher model was designed to integrate both panoramic radiographic images and structured clinical metadata, allowing it to learn complex associations between anatomical features and diagnostic labels. Training was performed using supervised learning techniques, with regularization strategies applied to enhance generalization.

Once the teacher model was trained, its output probabilities were used to guide the training of the student model. The student model replicated only the image-processing pathway of the teacher and was trained exclusively on radiographic input, without access to clinical metadata.

This allowed for a more lightweight architecture suitable for practical use in clinical environments, particularly when structured patient data may be unavailable. Through the distillation process, the student model was optimized to approximate the performance of the teacher while significantly reducing model complexity.

### 2.6.3 Model Selection

A preliminary evaluation phase was conducted to compare multiple convolutional neural network (CNN) architectures as potential teacher models for the classification of anatomical sectors in panoramic radiographs. The goal was to identify models capable of capturing the spatial and structural complexity of dental anatomy in 2D imaging. The selection process focused on architectural design, training behavior, and theoretical suitability for medical image classification tasks. The following CNN models were included in the comparison: ResNet50 [25], EfficientNetV2 [26], EfficientNetB0 [27], DenseNet121 [28].

### 2.6.4 Model Performance Evaluation

The performance of the classification models was assessed using a comprehensive set of evaluation metrics widely adopted in machine learning. Core classification metrics included accuracy, precision, and recall, with particular emphasis on weighted recall to account for class imbalance across the sector categories. To further quantify the discrepancy between predicted and actual class labels, regression-based metrics were also computed, including Mean Absolute Error (MAE), Mean Squared Error (MSE), and Root Mean Squared Error (RMSE). Formal definitions and formulas for all evaluation metrics are reported in *Appendix A*.

# 3.Results

### 3.1 Sample selection process

Of the 17,854 PRs that were analyzed, a total of 1,066 were included in the final dataset, resulting in 1,528 unerupted maxillary canines, according to the inclusion and exclusion criteria defined in the study protocol.

### 3.2 Analysis of the intra- and inter-observer reproducibility of the SCSs

The results from the training evaluation found an agreement between the examiners and the trainer ranging from moderate to almost perfect when using 5 and 4 classification systems, whereas no examiner showed lower than substantial agreement when using Kim's classification system, as shown in *Table 2*.

The agreement (k values) between observations (T0 versus T1) for each examiner ranged from 0.72 to 0.87 for 5-SCS; from 0.75 to 0.90 for 4-SCS; and from 0.80 to 0.92 for 3-SCS (*Table 3*), suggesting that, on average, the 3-SCS exhibits a greater intra-examiner reliability than 5 and 4 ones.

When comparing Os (examiners 1-3) and GDPs (examiners 4-6) in intra-examiner agreement, no differences were found for any SCS (*Table 4*), suggesting that the educational background did not affect the sector evaluation (p>0.05).

The overall inter-examiners agreement assessed using Fleiss' Kappa indicates a moderate to substantial level of consistency among all examiners at both T0 and T1 across the three classification systems (*Table 5*). On average, Kim's classification system yielded the highest overall agreement score, with a Fleiss' Kappa of 0.76 (95% CI: 0.74–0.79) at T0 and 0.69 (95% CI: 0.67–0.71) at T1. The clinical experience of the examiners did not affect the level of agreement when using Lindauer's (T0: p=0.73; T1: p=0.08), Kim's (T0: p=0.55;

T1: p=0.07) classification system. When using the 5-SCS, GDPs exhibited a higher level of agreement compared to Os at T0 (p=0.01), whereas the difference between the groups was not significant at T1 (p=0.05).

Overall, the 3-SCS introduced by Kim proved to be the method with the highest inter- and intra-observer reproducibility, and it is not influenced by the educational background.

## 3.3 Development of deep learning model for sector classification

### 3.3.1 Dataset Composition

The dataset that was used for deep learning model developing consisted in 1528 unerupted maxillary canines, classified according to the Kim's SCS. Based on the trainer analysis, the distribution of the samples across the three anatomical sectors was as follows: 592 samples were assigned to Sector A; 568 samples to Sectors B; 368 samples to Sectors C.

### 3.3.2 CNN model Selection

Among the evaluated CNN architectures, DenseNet121 achieved the highest overall accuracy of 76.4% in sector classification. In comparison, ResNet50 reached an accuracy of 67.2%, EfficientNetV2 achieved 58.4%, and EfficientNetB0 obtained 49.6%. Given its superior performance, DenseNet121 was chosen as the teacher model for the knowledge distillation.

### 3.3.3 Model Training

A two-step strategy based on DenseNet121 was adopted. First, a multimodal *teacher model* was trained using both panoramic radiographs and structured clinical data (canine depth, angular deviation, and root maturity). The image branch of the network used a DenseNet121 backbone pretrained on ImageNet, fine-tuned on our dataset. Then, a student model was trained using only radiographs, replicating the teacher's predictions via knowledge distillation (cross-entropy + divergence loss).

Both models were trained on the same dataset, consisting of 1,222 images (80%) for training and 306 images (20%) for validation, ensuring consistency in performance comparison (*Figure 2*).

**Figure 2;** Workflow of the AI model development using panoramic radiographs for unerupted maxillary canine detection, employing a knowledge distillation approach with DenseNet121.

### 3.3.4 Classification Results

The model was evaluated on a validation set of 306 PRs (20% of the full dataset). The student model, trained via knowledge distillation and relying exclusively on image input, achieved an overall classification accuracy. A per-class analysis revealed marked differences in sensitivity and precision across the three sectors.

- Class 1 (Sector A) was correctly 114 out of 118 cases (recall 96.61%, precision 71.25%), indicating high sensitivity but an overprediction tendency, likely due to class imbalance and visual similarity with adjacent sectors.
- Class 2 (Sectors B) showed high precision (87.50%), whereas recall fell to 49.56%, with many cases misclassified as Sector A, reflecting the anatomical ambiguity of this intermediate region.
- Class 3 (Sectors C) exhibited a precision of 76.19% and a recall of 85.33%, correctly classifying 64 out of 75 images.

A more detailed analysis of prediction errors is provided by the confusion matrix, which maps true class labels against predicted outputs. As shown in *Table 6*, Sector C showed the highest consistency, while Sector B was most frequently misclassified as Sector A.

### 3.3.5 Performance Metrics

Model performance was assessed through standard quantitative metrics. The Mean Absolute Error (0.258) and Mean Squared Error (0.304) indicated low deviation and infrequent severe errors, while the Root Mean Squared Error (0.551) confirmed close alignment between predictions and true labels. The model achieved an overall accuracy of 76.47% and a recall of 77.17%, demonstrating robust and sensitive classification performance based solely on radiographic input.

## 4. Discussion

The present study results indicated substantial reproducibility from all three classification systems based on kappa statistics, with precision ranging from moderate to almost perfect agreement. Considering the above, in Erikson and Kurol's [13] and Lindauer's [22] SCS, the majority of inaccuracies occurred when determining neighboring sectors, the union of categories in Kim's classification [23] highlighted greater agreement and enhanced its degree of reproducibility. Furthermore, reliability was not influenced by educational background (*Table 2*). Similar results were shown in the inter-examiner reliability, with the K values rated as 'substantial agreement' or 'almost perfect agreement'. Higher agreement was found in the 3-SCS, and the educational background of the examiners did not influence these results (*Table 4*). Therefore, Kim's SCS was used to train the AI model.

Among the evaluated CNN architectures, DenseNet121 outperformed all other models, achieving the highest accuracy and demonstrating a consistent ability to capture fine-grained anatomical differences across sectors. Coherently, DenseNet-121 has proved to be great for medical imaging applications because of its high level of connectivity that makes it efficient in image detection and classification tasks. Notably, in dental radiology, Lee et al. demonstrated that DenseNet-121 could classify dental caries with a level of accuracy comparable to that of expert dentists, achieving 95.7% accuracy in the detection of early dentinal lesions [29]. Moreover, DenseNet-121 achieved outstanding accuracy of 99.5% in classifying common dental pathologies such as periapical periodontitis, caries and cysts on digital dental PRs, surpassing other CNN models [30].

In the present study, the full deep learning model demonstrated strong overall classification accuracy (76.47%). This indicates its potential applicability in automated diagnostic workflows for maxillary canine impaction. However, the results highlighted notable differences in sensitivity and precision among the classes (*Table 6*).

Sector A exhibited the highest recall (96.61%) indicating excellent sensitivity but a lower precision (71.25%) due to a relatively high number of false positives, which mostly consisted of misclassifications from Sector B. This suggests a slight bias of the model, probably induced by the possible class imbalance and the similarity of features in the adjacent sectors.

Sector B was the most difficult case to classify for the model, with a high precision (87.50%) but low recall (49.56%). On PRs, Sectors A and B are separated only by the long axis of the lateral incisor and their boundary may appear less distinct due to superimposition, image distortion, and variable canine inclination. Hence, the crown of canine in the middle area of the lateral incisor might resemble both sectors and both human observers and the model might give different classifications. From a clinical perspective this misclassification is relevant as the canines in Sector B generally represent the cases with the highest risk of impaction and root resorption of the lateral incisor, which require interceptive treatment at an early age (i.e. extraction of the primary canine or orthodontic traction), whereas the canines in Sector A usually have a more favorable eruption pattern and can be considered for observation only. Thus, an error in the model that overestimates Sector B might lead to unnecessary treatment while the underestimation of Sector A might delay the appropriate treatment and thus the need for a correct and reliable classification is evident. This may mean that enhanced model attention mechanisms or the inclusion of auxiliary clinical features beyond image input are needed.

Sector C showed the most balanced performance, achieving a precision rate of 76.19% and a recall rate of 85.33%. The higher classification consistency suggests that the features that define Sector C are more distinctive and easier for the model to recognize. This is further confirmed by the confusion matrix, which shows that Sector C had the fewest misclassifications and strong alignment between the predicted and true labels.

Overall, the model's accuracy was comparable to general dental practitioners and slightly lower than that of orthodontists. This tool could be compared to human decision, and if the sector classification information is integrated with more data as alpha angle, the depth of inclusion, the presence or absence of lateral incisors adjacent to the impacted canine, dental anomalies of the upper incisors, and the presence of a neoplasm (cyst, odontoma) in the upper jaw, then the model's accuracy could increase and be more.

The results of this study are valid within the following limits. First, a possible selection bias must be considered, as the exclusion of radiographs with impacted canines and agenesis of the premolars and/or the presence of cysts led to the training of a model that was limited in its recognition of radiographic images. The second limitation was the modest number of radiographs, especially for Sectors B and Sector C. Moreover, external validation is required before clinical application, as the model was trained on a single-institution dataset.

Future developments should therefore focus on expanding the dataset, particularly regarding the underrepresented sectors, in order to strengthen the model's ability to distinguish between all sectors. Increasing the number and type of samples with X-rays showing agenesis and cysts would mitigate class imbalance, improve robustness and reduce overfitting towards dominant patterns.

# 5. Conclusion

The sector classification system provides a reproducible and objective method for determining the position of impacted maxillary canine and has been shown not to be influenced by the education background. Furthermore, the AI model demonstrated strong overall classification accuracy, supporting its potential as a valuable diagnostic tool. Notably, the model was particularly effective in identifying canines with a higher likelihood of impaction (Sector A) compared to those with a more favorable eruption trajectory (Sector C). This study represents a first step towards fully automated PRs analysis for canine impaction classification.

**Tables**

**Table 1.** Sector classification systems.

| 5-sector classification system [13] | S1: the area distal to a line tangent to the distal heights of the contour of the lateral incisor crown and root |
|---|---|
| | S2: was the area between a line tangent to the distal heights of the contour of the lateral incisor crown and root and a line bisecting the mesiodistal dimension of the lateral incisor along the long axis of the tooth |
| | S3: the area between a line bisecting the mesiodistal dimension of the lateral incisor along the long axis of the tooth and a line tangent to the mesial heights of the contour of the lateral incisor crown and root |
| | S4: the area between a line tangent to the mesial heights of the contour of the lateral incisor crown and root and a line bisecting the mesiodistal dimension of the central incisor along the long axis of the tooth |
| | S5: the area mesial to a line bisecting the mesiodistal dimension of the central incisor along the long axis of the tooth |
| 4-sector classification system [22] | I: the area distal to a line tangent to the distal heights of the contour of the lateral incisor crown and root |
| | II: the area between a line tangent to the distal heights of the contour of the lateral incisor crown and root and a line bisecting the mesiodistal dimension of the lateral incisor along the long axis of the tooth |
| | III: the area between a line bisecting the mesiodistal dimension of the lateral incisor along the long axis of the tooth and a line tangent to the mesial heights of the contour of the lateral incisor crown and root |
| | IV: the area mesial to a line tangent to the mesial heights of the contour of the lateral incisor crown and root |

| | 3-sector classification system [23] | A: the area distal to a line tangent to the distal heights of the contour of the lateral incisor crown and root |
|---|---|---|
| | | B: the area between a line tangent to the distal heights of the contour of the lateral incisor crown and root and a line tangent to the mesial heights of the contour of the lateral incisor crown and root |
| | | C: the area mesial to a line tangent to the mesial heights of the contour of the lateral incisor crown and root |

**Table 2.** Training evaluation process: agreement (Cohen's K) between trainer and examiner in sector classification with the three different classification systems.

| | **K value (95% CI)** | | | | | |
|---|---|---|---|---|---|---|
| | T0 | | | T1 | | |
| **Examiner** | Erikson and Kurol | Lindauer | Kim | Erikson and Kurol | Lindauer | Kim |
| **1** | 0.76 (0.70-0.82) | 0.78 (0.72-0.84) | 0.81 (0.76-0.87) | 0.70 (0.63-0.76) | 0.70 (0.64-0.76) | 0.75 (0.69-0.82) |
| **2** | 0.81 (0.76-0.86) | 0.84 (0.79-0.89) | 0.87 (0.81-0.91) | 0.81 (0.76-0.86) | 0.82 (0.77-0.88) | 0.86 (0.81-0.91) |
| **3** | 0.73 (0.63-0.76) | 0.76 (0.71-0.82) | 0.79 (0.73-0.84) | 0.65 (0.58-0.71) | 0.66 (0.59-0.73) | 0.74 (0.68-0.81) |
| **4** | 0.59 (0.52-0.66) | 0.62 (0.56-0.69) | 0.69 (0.62-0.76) | 0.54 (0.47-0.60) | 0.60 (0.53-0.66) | 0.63 (0.56-0.70) |
| **5** | 0.65 (0.59-0.72) | 0.69 (0.62-0.75) | 0.73 (0.66-0.80) | 0.59 (0.52-0.66) | 0.62 (0.55-0.68) | 0.69 (0.62-0.76) |
| **6** | 0.64 (0.57-0.70) | 0.78 (0.72-0.84) | 0.82 (0.76-0.87) | 0.69 (0.62-0.75) | 0.77 (0.71-0.82) | 0.82 (0.77-0.88) |

*Abbreviations: CI, confidence interval; T0, time of examiners' first analysis of radiograph; T1, time of second examiners 'analysis of radiographs.*

**Table 3.** Intra-observer (T0 versus T1) agreement for each examiner assessed through Cohen's K.

| | **K value (95% CI)** | | |
|---|---|---|---|
| **Examiner** | **Erikson and Kurol** | **Lindauer** | **Kim** |
| 1 | 0.79 (0.73-0.84) | 0.78 (0.72-0.83) | 0.81 (0.75-0.86) |
| 2 | 0.83 (0.78-0.88) | 0.86 (0.81-0.91) | 0.87 (0.83-0.92) |
| 3 | 0.76 (0.70-0.82) | 0.77 (0.71-0.83) | 0.81 (0.75-0.87) |
| 4 | 0.84 (0.79-0.89) | 0.87 (0.862-0.91) | 0.90 (0.86-0.94) |
| 5 | 0.72 (0.66-0.78) | 0.75 (0.69-0.81) | 0.80 (0.74-0.86) |
| 6 | 0.87 (0.83-0.92) | 0.90 (0.86-0.94) | 0.92 (0.88-0.96) |

*Abbreviations: CI, confidence interval; T0, time of examiners' first analysis of radiograph; T1, time of second examiners 'analysis of radiographs.*

**Table 4.** Intra-observer (T0 versus T1) agreement (Cohen's K) for groups of examiners (general dental practitioners versus orthodontists) and overall. Between-groups differences assessed through z-Test.

| | **Erikson and Kurol** | **Lindauer** | **Kim** |
|---|---|---|---|

|       | K value (95% CI) | p-value | K value (95% CI) | p-value | K value (95% CI) | p-value |
|-------|------------------|---------|------------------|---------|------------------|---------|
| Os    | 0.79 (0.76-0.82) | .34     | 0.80 (0.77-0.83) | .07     | 0.83 (0.80-0.86) | .07     |
| GDPs  | 0.81 (0.78-0.84) |         | 0.84 (0.81-0.87) |         | 0.85 (0.83-0.90) |         |
| Overall | 0.80 (0.78-0.82) |       | 0.82 (0.80-0.84) |         | 0.85 (0.83-0.87) |         |

*Abbreviations: CI, confidence interval; GDPs, general dental practitioners; Os, orthodontists; p, p-value; T0, time of examiners' first analysis of radiograph; T1, time of second examiners 'analysis of radiographs.*

**Table 5.** Inter-observer agreement (Fleiss' K) for groups of examiners (general dental practitioners versus orthodontists) and overall. Between-groups differences assessed through z-Test.

| Examiner | T0 | | | | | | T1 | | | | | |
|----------|----|----|----|----|----|----|----|----|----|----|----|----|
|          | *Erikson and Kurol* | | *Lindauer* | | *Kim* | | *Erikson and Kurol* | | *Lindauer* | | *Kim* | |
|          | K value (95% CI) | *p* | K value (95% CI) | *p* | K value (95% CI) | *p* | K value (95% CI) | *p* | K value (95% CI) | *p* | K value (95% CI) | *p* |
| Os       | 0.72 (0.68-0.75) | .01* | 0.75 (0.71-0.79) | .73 | 0.78 (0.71-0.83) | .55 | 0.66 (0.63-0.70) | .05 | 0.69 (0.65-0.73) | .08 | 0.74 (0.70-0.79) | .07 |
| GDPs     | 0.78 (0.74-0.81) |      | 0.74 (0.70-0.78) |     | 0.76 (0.71-0.80) |     | 0.61 (0.58-0.65) |     | 0.64 (0.60-0.68) |     | 0.68 (0.63-0.73) |     |
| Overall  | 0.69 (0.68-0.71) |      | 0.73 (0.71-0.74) |     | 0.76 (0.74-0.79) |     | 0.60 (0.59-0.62) |     | 0.64 (0.63-0.66) |     | 0.69 (0.67-0.71) |     |

*Abbreviations: CI, confidence interval; GDPs, general dental practitioners; Os, orthodontists; p, p-value; T0, time of examiners' first analysis of radiograph; T1, time of second examiners 'analysis of radiographs.*

**Table 6.** Confusion matrix

| True class | Predicted: Class 1 | Predicted: Class 2 | Predicted: Class 3 |
|------------|--------------------|--------------------|--------------------|
| Class 1 (Sector A) | 114 | 2  | 2  |
| Class 2 (Sector B) | 41  | 56 | 16 |
| Class 3 (Sector C) | 5   | 6  | 64 |

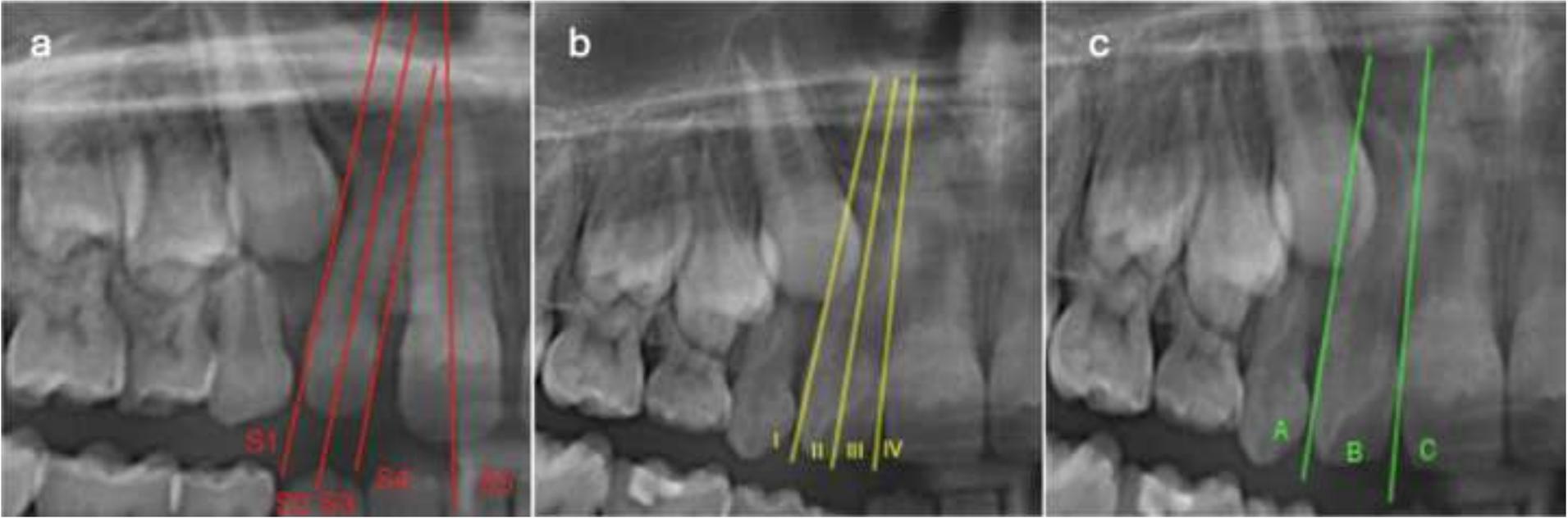

Figure 2 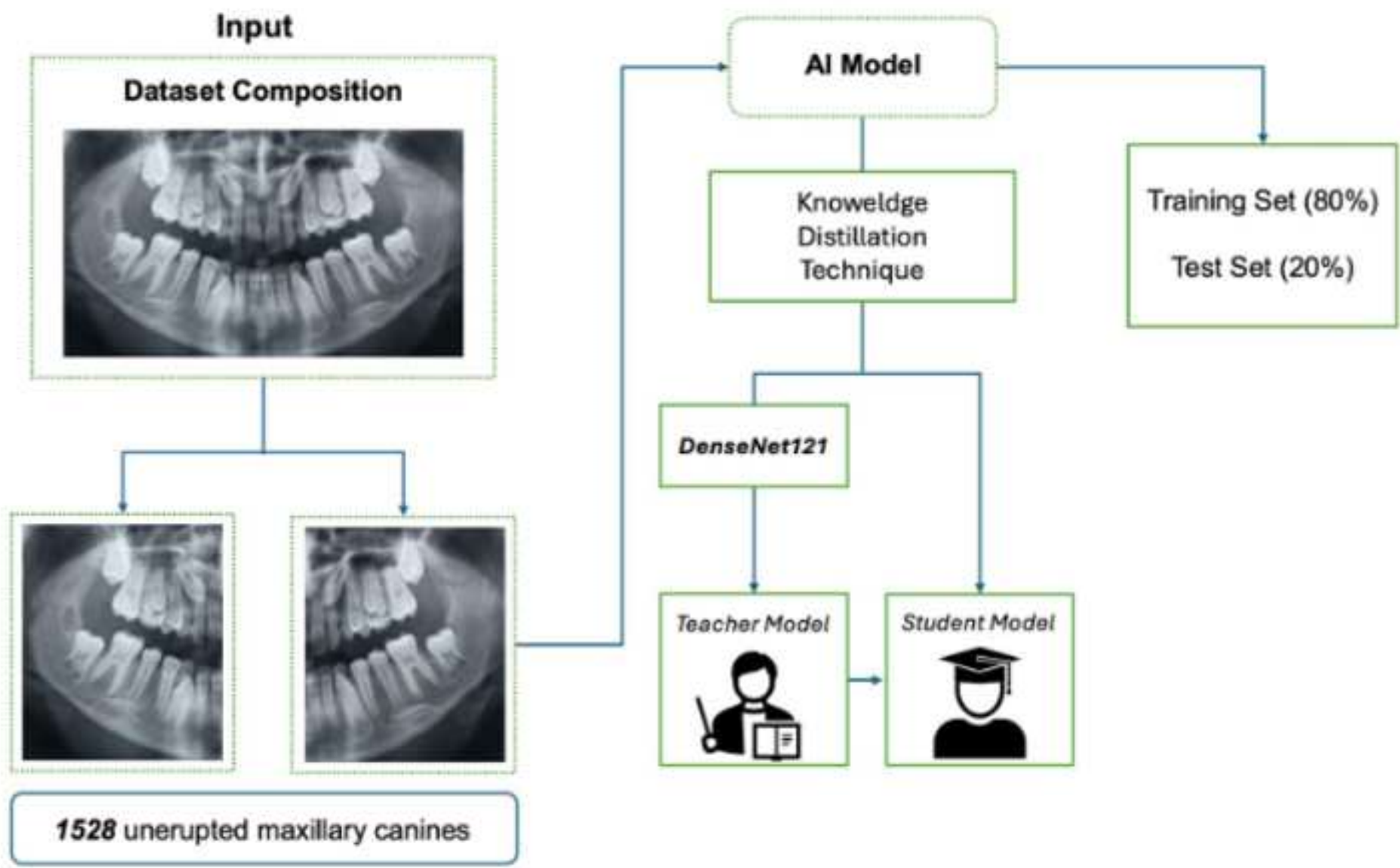

# APPENDIX

**Evaluation Protocol**

Model performance was evaluated using standard metrics derived from classification outcomes (true positives, false positives, true negatives, false negatives). The following metrics were computed:

- **Accuracy (ACC)**
  
  Proportion of correct predictions over the total cases:

$$\text{ACC} = \frac{TP + TN}{TP + TN + FP + FN}$$

- **Recall (Sensitivity)**
  
  Ability to correctly identify all true cases of each class:

$$\text{Recall}_i = \frac{TP_i}{TP_i + FN_i}$$

  A weighted version was calculated to account for class imbalance:

$$\text{Weighted Recall} = \frac{\sum_i (\text{Recall}_i \cdot \text{Support}_i)}{\sum_i \text{Support}_i}$$

- **Precision**
  
  Proportion of correct positive predictions:

$$\text{Precision}_i = \frac{TP_i}{TP_i + FP_i}$$

- **Mean Absolute Error (MAE)**
  
  Average absolute difference between prediction and reality:

$$\text{MAE} = \frac{1}{n} \sum_{i=1}^{n} |y_i - \hat{y}_i|$$

- **Mean Squared Error (MSE)** and **Root Mean Squared Error (RMSE)**
  Indicators of overall prediction error and its magnitude:

$$\text{MSE} = \frac{1}{n}\sum_{i=1}^{n}(y_i - \hat{y}_i)^2, \quad \text{RMSE} = \sqrt{\text{MSE}}$$

These metrics provided a comprehensive picture of how well the models performed, both in terms of classification accuracy and prediction reliability.

**Authors Contribution Statement**

Conceptualization, M.G. and S.M.; methodology, D.C. and S.M.; software, D.R. and L.R.; validation, S.M.; formal analysis, F.C. and M.G.; investigation, D.C. and M.G.; resources, S.M. and T.B.; data curation, D.C., F.C., L.R.; writing— original draft preparation, F.C. and M.G.; writing—review and editing, D.C. and T.B.; visualization, D.R.; supervision, S.M.; project administration, S.M.